\documentclass[runningheads,a4paper]{llncs}

\usepackage{amsmath}
\usepackage{amssymb}
\usepackage{graphicx}
\usepackage{wrapfig}
\usepackage{multirow}
\usepackage{array}
\usepackage{xcolor}
\usepackage{balance} 
\usepackage[absolute]{textpos} 
\usepackage[ruled,vlined]{algorithm2e}
\usepackage{epstopdf}
\usepackage{enumitem} 

\newcolumntype{C}[1]{>{\centering\let\newline\\\arraybackslash\hspace{0pt}}m{#1}}

\newcommand\T{\rule{0pt}{3.4mm}}
\newcommand\B{\rule[-1.5mm]{0pt}{0pt}}

\usepackage{url}
\newcommand{\keywords}[1]{\par\addvspace\baselineskip
\noindent\keywordname\enspace\ignorespaces#1}

\def\unimelb{\textsuperscript{\dag}}
\def\sutd{\textsuperscript{\ddag}}

\begin{document}

\mainmatter  

\title{Mining Influentials and their Bot Activities on Twitter Campaigns}

\titlerunning{Mining Influentials and their Bot Activities on Twitter Campaigns}

\toctitle{Mining Influentials and their Bot Activities on Twitter Campaigns}

%
%
\author{Shanika Karunasekera\unimelb \and Kwan Hui Lim\unimelb\sutd \and Aaron Harwood\unimelb}
\authorrunning{S. Karunasekera, K. H. Lim, A. Harwood} 
\tocauthor{Shanika Karunasekera, Kwan Hui Lim, Aaron Harwood} 
\institute{
\unimelb The University of Melbourne and \sutd Singapore University of Technology and Design\\
\email{\{karus,kwan.lim,aharwood\}@unimelb.edu.au}
}

%
%

\maketitle

\begin{abstract}
Twitter is increasingly used for political, advertising and marketing campaigns, where the main aim is to influence users to support specific causes, individuals or groups. We propose a novel methodology for mining and analyzing Twitter campaigns, which includes: (i) collecting tweets and detecting topics relating to a campaign; (ii) mining important campaign topics using scientometrics measures; (iii) modelling user interests using hashtags and topical entropy; (iv) identifying influential users using an adapted PageRank score; and (v) various metrics and visualization techniques for identifying bot-like activities. While this methodology is generalizable to multiple campaign types, we demonstrate its effectiveness on the 2017 German federal election. 
\end{abstract}

\section{Introduction}

Twitter is a popular microblogging service that is also increasingly being used as a platform for political, advertising and marketing campaigns~\cite{Tumasjan-icwsm10,Prasetyo-ht15,lim-wias16}. Twitter was the dominant platform for breaking news on the 2016 US Presidential elections, generating more than a billion tweets in the lead-up to the elections and 40 million tweets during election day itself~\cite{nyt-election16}. This widespread use of Twitter as a campaign platform demonstrates the immense impact that Twitter has on our society, and has sparked research interest in data-driven political science. 
The primary research question we aim to address in our work is which influential users used bot-based dissemination strategies in Twitter campaigns through discovering the influentials, rather than targeting specific users such as party leaders or presidential candidates, identified a priori.

\vspace{2mm}
{\noindent\bf Related Work}.  
Election results have important societal implications and has garnered the interest of the academic community~\cite{Tumasjan-icwsm10,Prasetyo-ht15}, who attempt to predict the winning party of elections.
Closely related to election prediction are works that aim to apply classification or prediction models to predict the political preferences or alignment of individual users~\cite{Conover-passat11,Boutet-icwsm12}. 
Similarly, another research area closely related to the study of Twitter campaigns is the detection of bots on Twitter, which is well-studied in recent years~\cite{Davis-www16,Varol-icwsm17}. 
While these earlier works study interesting aspects of Twitter research, they focus on smaller sub-problems, such as election prediction, individual classification and bot detection, without understanding the entire campaign as a holistic process.

In contrast, we develop a methodology for studying entire Twitter campaigns, drawing from and combining unsupervised techniques from topic modelling and network analysis to better understand the mechanics of Twitter campaigns. We believe that our proposed methodology will be repeatable and applicable in a range of domains due to the use of unsupervised learning techniques and measurements, compared to works that utilize prediction/classification algorithms, which potentially suffer from the lack of quality ground truth for training, overfitting of data, obsoleteness of classification features and being domain-specific.

\vspace{1mm}
{\noindent\bf Contributions}. 
Our main contribution include developing a novel methodology (\S\ref{sectMethodology}) for analyzing Twitter campaigns, in terms of the discussion topics, campaign influencers, audiences, interactions and bot activities, which comprises the following capabilities: (i) An approach for identifying and filtering the most relevant and important topics in the campaign, adapting from centrality and density measures used in scientometrics analysis (\S\ref{sectIdenTopTopics}); (ii) A model of user interest and topical interest variance based on a hashtag-based interest measure and topical entropy measure (\S\ref{sectUserTopic}); (iii) An influence score for identifying important Twitter users, based on a variant of the PageRank algorithm applied on a retweet network (\S\ref{sectInfluenUser}); and (iv) Various measures for identifying users that potentially employ bots, and visualization techniques for identifying the bots themselves. (\S\ref{sectBotActivities}).
We also demonstrate this methodology on the 2017 German federal election (comprising 8.88 million tweets) and discuss our main findings in terms of the discussion topics, influential users, retweeters and bot-like behaviour (\S\ref{sectResults}).

\section{Methodology for Twitter Campaign Analysis}
\label{sectMethodology}

We now elaborate on our methodology for analyzing Twitter campaigns.

\subsection{Data Collection and General Campaign Topic Detection}

We first collect campaign-related tweets using the Real-time Analytics Platform for Interactive Data Mining (RAPID)~\cite{lim-ecml18}, which provides an interface to the Twitter API. Thereafter, we proceed to detect the topics frequently discussed by users in the campaign and identify the important topics among the larger set of detected topics. 
For modelling the topics discussed by users in the campaign, we utilize a clustering-based approach on a hashtag co-occurence graph~\cite{lim-bigdata17}, which constructs a hashtag graph (based on hashtag co-occurences in tweets), then applies the Louvain Algorithm~\cite{blondel-jsm08} on this graph to detect campaign topics. Based on these topics, we next filter a subset that are most relevant to the campaign.

\subsection{Identifying and Filtering Relevant Topics}
\label{sectIdenTopTopics}

We now identify a subset of topics that are most relevant to the campaign, among all topics. For this purpose, we select the top $N$ topics that have at least $N$ hashtags (similar to H-index used for identifying significant publications of a researcher~\cite{Hirsch-pnas05}); this technique does not require setting thresholds on the number of topics and hashtags, and was able to filter the significant topics as we demonstrate in our results.
 
We then proceed to identify the topics that are central and relevant to the campaign, from the significant topics. For this purpose, we adapt a popular technique for topic selection from scientometrics analysis, the strategic diagram~\cite{An-sci11}. The strategic diagram is a two-dimensional scatter plot of topic density vs centrality, divided into four quadrants based on the mean or median centrality and density. Topics in the first quadrant (top right), and fourth quadrant (bottom right), which are topics with high centrality are considered most relevant to the theme, hence we select topics that are in the first and fourth quadrants. Using similar definitions as~\cite{An-sci11}, the centrality $Cen(c)$ and density $Den(c)$ of a topic cluster $c$ are defined as: 
\begin{equation}
Cen(c) = \frac{\sum\limits_{h_i \in H^c}~\sum\limits_{h_j \in H - H^c}~e_{{h_i},{h_j}}}{\big(|H| - |H^c|\big) |H^c|}
\label{centrality}
\end{equation}
\begin{equation}
Den(c) = \frac{\sum\limits_{h_i \in H^c}~\sum\limits_{h_j \in H^c, h_i \ne h_j}~e_{{h_i},{h_j}}}{|H^c|-1}
\label{density}
\end{equation}
 
In the above, a topic cluster $c$ is represented by an undirected graph $G^c = (H^c, E^c)$, where $H^c \subset H$ is the set of hashtags in the topic cluster (and $H$ is the set of all hashtags) and $E^c_{{h_i},{h_j}} \subset E$ is the set of edges (where $e_{{h_i},{h_j}}=1$ if hashtags $h_i$ and $h_j$ are used together in any tweet, and $e_{{h_i},{h_j}}=0$ otherwise).

\subsection{Modelling and Understanding Users' Topics of Interest}
\label{sectUserTopic}

We now progress from understanding the general campaign topics to modelling the unique topical interests of each user, which we describe in the next section.

\vspace{-3.0425mm}
\subsubsection{User Topical Interests}
Based on the detected topics in the campaign, 
we proceed to model the level of user interests in each of these topics. We represent the topical interests of a user $u$ as a vector $\vec{Int^u} = \langle int^u_{t1},...,int^u_{tn} \rangle$, where $int^u_{t} = [0,1]$ denotes the user's interest level in topic $t$ as a continuous value from 0 (not interested) to 1 (very interested).
Given that $c_t = \{h_1,...,h_m\}$ is the topic cluster of representative hashtags for topic $t$ 
and $H_u$ denotes all hashtags posted by user $u$, we define the interest level of user $u$ in topic $t$ as:
\begin{equation}
int^u_t = \frac{1}{|H_u|} \sum_{h \in H_u} tf(h,c_t)
\label{topicInterest}
\end{equation} 

{\noindent where term frequency $tf(h,c_t)$ denotes the number of times user $u$ uses a hashtag $h$ that belongs to topic cluster $c_t$. That is, we measure a user's interest level in topic $t$ based on the number of times he/she used a hashtag that belongs to a topic, relative to his/her total number of hashtags used. }

\vspace{-3.0425mm}
\subsubsection{Topical Interest Entropy}
Apart from modelling users' topical interests, we are also interested in how varied or focused their topical interests are, i.e., do they have specialized and high interest in a small set of topics, or a general but low interest in a wide set of topics? 
To determine the diversity of a user's topical interest, we adapted the measure of topic entropy, which has been used to study the focus areas of conferences on different research areas~\cite{Hall-emnlp08}. 

Given a set of topics $T$, we define topic entropy for a user $u$ as:
\begin{equation}
Ent(u) = -\sum_{t \in T} int^u_t~\mathrm{log}~int^u_t
\label{entropy}
\end{equation} 

{\noindent where $int^u_t = [0,1]$ denotes the the interest level of user $u$ in topic $t$ (see Eqn.~\ref{topicInterest}).}

\subsection{Identifying Influential Users}
\label{sectInfluenUser}

A key aspect of any Twitter campaign is to identify a set of influential users that are crucial in influencing the activities and opinions of other general users~\cite{Riquelme-ipm16,Xiao-jwe14}. We next introduce our definition of a user influence score and describe its usage in identifying influential users and understanding their representative topics.

\vspace{-3.0425mm}
\subsubsection{User Influence Score}
In Twitter campaigns, influential users tend to be consistently and frequently retweeted, and similarly for the Internet (academia), influential websites (authors) are highly referenced by other websites (authors). Thus we adopted the PageRank algorithm~\cite{Brin-cn12} for measuring a Twitter user's influence level $Inf(u)$. PageRank is traditionally used to determine the importance of websites based on their incoming links (from other websites). Although PageRank and its variants have been used in the literature to identify influence on Twitter, the focus has been on the follower network~\cite{Kwak-www10} and with adaptations to take into consideration topics and interactions~\cite{Weng-wsdm10}. Instead, we compute influence based only on the retweet network because our goal is to identify influentials in terms of information dissemination during the period of interest. 

For each day $d$, we construct a retweet graph that comprises a set of users $U^d$ who have retweeted a tweet on day $d$ and $RT^d(u_o, u)$ denotes the number of times user $u_o$ retweeted user $u$ on day $d$. Based on these definitions, we calculate the PageRank score $PR^d(u)$ of user $u$ on day $d$ as:
\begin{equation}
PR^d(u) = \frac{(1-\beta)}{|U^d|} + \beta \sum_{u_o \in U^d} \frac{PR^d(u_o)}{RT^d(u_o,u)}
\label{pagerank}
\end{equation} 

{\noindent where we set $\beta = 0.85$ in accordance to the best value determined in the original PageRank paper~\cite{Brin-cn12} and for Twitter related works~\cite{Weng-wsdm10}.}

In turn, the normalized PageRank score $NPR^d(u)$ of a user $u$ for day $d$ is:
\begin{equation}
NPR^d(u) = \frac{PR^d(u)}{\max\limits_{u_i \in U^d} PR^d(u_i)}
\label{normpagerank}
\end{equation} 

That is, a user's normalized PageRank score $NPR^d(u)$ on day $d$ is based on his PageRank score $PR^d(u)$ divided by the max PageRank scores on day $d$.

Let $N$ be the tweeting duration (in days), the influence level of user $u$ is:
\begin{equation}
Inf(u) = \frac{1}{N} \sum_{d \in N} NPR^d(u) 
\label{userInfluence}
\end{equation} 

{\noindent where $NPR^d(u)$ denotes the normalized PageRank score of a user $u$ on day $d$ that is based on the PageRank score $PR^d(u)$ normalized by the max PageRank score on day $d$ (Equations~\ref{pagerank} and~\ref{normpagerank}). 
We average the normalized PageRank score $NPR^d(u)$ by $N$ days to identify users who are consistenly influential, i.e., to avoid users who are mis-identified as influential over the period because of high influence on a particular day due to a hot topic.}

\subsection{Identifying Bot Activities}
\label{sectBotActivities}

To identify if an influencer is employing bots for retweeting, we need to perform two tasks: (i) first, identify users who are employing such a strategy; (ii) second, identify which of a influencer's retweeters are bots. We propose various measures and techniques for these tasks, which we discuss next.

\vspace{-3.0425mm}
\subsubsection{Unique Retweet Ratio}

For the first task of identifying users employing bots, we propose the Unique Retweet Ratio $UR^p(u)$ based on the number of unique users retweeting an influential user's tweets. 
Given a retweet graph $G_{RT} = (U, RT)$ where $U$ denotes the set of users and $RT$ the set of edges such that $RT^p(u_o, u)$ is the number of times user $u_o$ retweeted user $u$ during a period $p$, we calculate the unique retweet ratio $UR^p(u)$ of user $u$ at time period $p$ as:
\begin{equation}
UR^p(u) = \frac{|U^p|}{\sum\limits_{u_o \in U}RT^p(u_o, u)}
\label{uniqueRetweet}
\end{equation} 

{\noindent where $U^p$ is the set of users who retweeted user $u$ during period $p$, i.e., $RT^p(u_o, u) > 0$. In short, we measure the unique retweet ratio $UR^p(u)$ of user $u$ at time period $p$ based on the number of unique retweeters, relative to the total number of retweets posted. The basic intuition is that a lower unique retweet ratio indicates a higher probability of a user employing bots for retweeting (hence a small number of unique users account for a large number of retweets).

\vspace{-3.0425mm}
\subsubsection{Direct Retweeter Influence}

We measure a user $u$'s direct retweeter influence $DI^p(u)$ as the transfer entropy from this user $u$ to his/her retweeters $r$~\cite{Steeg-www12}. To compute $DI^p(u)$ for the period we divide the period into equal size bins of time $T_x$. We then derive a binary vector for the user with a bin value of 1 if the user tweeted during the time period and 0 otherwise. We use this vector to compute the transfer entropy, as follows:
\begin{equation}
DI^p(u) = H(u_t|u^{(t-k)}_{t-1}) - H(u_t|u^{(t-k)}_{t-1},r^{(t-l)}_{t-1})
\label{transferEntropy}
\end{equation}

{\noindent where $H(u_t|u^{(t-k)}_{t-1})$ denotes our uncertainty regarding $u_t$ when only given the tweeting history of user $u$, while $H(u_t|u^{(t-k)}_{t-1},r^{(t-l)}_{t-1})$ denotes the reduction of uncertainty when a retweeters history $r_t$ is provided. In our application scenario, a high value of transfer entropy indicates that the user's retweet action was directly influenced by the original tweet.  Our intuition is that users that mechanically retweet the user of interest, hence bot-like behaviour, will have a high value of transfer entropy compared to users that retweet organically~\cite{Steeg-www12}.}

\vspace{-3.0425mm}
\subsubsection{Tweeting Volume of Retweeters}

For the second task, we want to identify which of a influencer's retweeters are exhibiting bot-like behaviours. Our initial approach is to analyze the average daily tweet volumes of users who retweeted the campaign influencers, with the intuition that bots have high tweet counts.

\vspace{-3.0425mm}
\subsubsection{Visualizing Influential Users Employing Bots}

To augment this second task, we propose a visualization technique that clearly highlights influentials that are employing bots and the characteristics of these tweets. In this visualization, we plot a scatterplot of retweeters (data points) for each influential user, where each retweeters colour, shape, and size denotes their excessive tweeting volume, number of influencers retweeted and influence level, respectively.

\vspace{-3.0425mm}
\subsubsection{Common Retweeters across Campaign Influencers}

To better measure the extent of common retweeters across the influential users, we propose a Retweet Jaccard measure, which is based on a pair-wise Jaccard similarity measure between a user of interest with other influential users. 
For a user $u$, his/her set of retweeters $R_u$ and other influential users $U^I$, we define his/her Retweet Jaccard $RJ(u)$ as:
\begin{equation}
RJ(u) = \frac{1}{|U^I|}\sum_{u^i \in U^I} \frac{|R_u \cap R_{u^i}|}{|R_u \cup R_{u^i}|}
\label{retweetJaccard}
\end{equation} 

{\noindent That is, we calculate the Retweet Jaccard $RJ(u)$ of a user $u$, based on the mean of his/her pair-wise Jaccard similarity with the remaining influential users. A high $RJ(u)$ value indicates that user $u$ shares many common retweeter than the other influential users, and a low $RJ(u)$ value indicates otherwise.}

\section{Results: A Case-study on the 2017 German federal election}
\label{sectResults}

We next describe the application of our methodology (\S\ref{sectMethodology}) on a case study of the 2017 German federal election, and highlight some of our main findings.

\subsection{Campaign Dataset}
\label{sectDatasetCollected}

We tracked and retrieved a total of 8.88 million tweets (generated by 629k users) related to the 2017 German federal election.
In our subsequent analysis, we studied the tweets based on two time periods (Period 1 from 01 to 31 Aug 17, and Period 2 from 01 to 23 Sep 17) 
to better understand the change in tweeting patterns, influential users and bot followers. For Period 1 we collected 4.09 million tweets from 382k users, and for Period 2 we collected 4.79 million tweets from 433k users. Also by applying the methodology for two distinct periods we were also able to demonstrate the repeatability of the methodology.

\subsection{Top 10 Campaign Topics}
Following which, we then detect the topics discussed by all users in terms of the two month preceeding the election. 
A qualitative analysis of this topic modelling shows that the clustered hashtags are good indications of the underlying topics. Our first observation is that, in spite of targeted tracking, the set of topics is diverse, some topics more related to the election, while other topics are less related (transient and peripheral).  The first three topics in each period appear to be more closely related to the election, containing: major election hashtags (\#btw, \#btw2017, \#btw17),
hashtags related to the parties (\#afd, \#cdu, \#spd, \#fdp, \#linke, \#csu,
\#grne) and
politicians (\#merkel, \#schulz, \#weidel). Topic T1 (both periods) relates to the Alternative for Germany political party, with representative hashtags of the party (\#afd, \#noafd), their politicians (\#gauland,\#weidel) and party slogan (\#traudichdeutschland), while Topic T10 relates to international relations/issues based on hashtags of various countries (Period 1) and the Ecological Democratic Party or \"Okologisch-Demokratische Partei (Period 2). From this larger list of topics, our next task is to identify a subset of topics that are directly related to the political campaign

\subsection{Filtering of Important Campaign Topics}

\begin{wrapfigure}{R}{0.35\columnwidth}
\vspace{-13mm}
  	\centering
    \includegraphics[width=0.34523\columnwidth, keepaspectratio=true, angle=0, trim={3mm 0mm 0mm 0mm}]{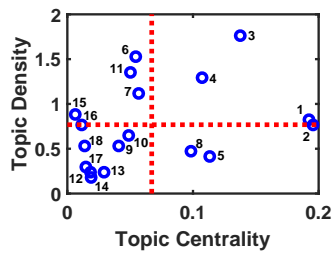}
\vspace{-2mm}
    \caption{Topic Density VS Topic Centrality for Period 1}
    \label{topicDenCen}
\vspace{-8mm}
\end{wrapfigure}

From the topic clusters we first select the $N$ significant clusters as described in Section 3.3. For Period 1, $N = 18$ and for Period 2, $N = 22$. Fig.~\ref{topicDenCen} shows a example plot of Topic Density $Den(c)$ VS Topic Centrality $Cen(c)$ (Equations~\ref{density} and~\ref{centrality} from \S\ref{sectIdenTopTopics}) for all detected topic clusters in Period 1. 
The red vertical line shows the mean centrality and the red horizontal line shows the mean density. 
We select topics from quadrants 1 (top right) and 4 (bottom right): Topics 1, 2, 3, 4, 5 and 8 for Period 1, and Topics 1, 2, 3, 4, 5 and 7 for Period 2 as the most relevant topics for further analysis. We also observe that the selected topics covered key events and topics relating to the election.

\subsection{User Interests and Topical Entropy}

Using our model of user topical interest and topical entropy (introduced in \S\ref{sectUserTopic}), we now study the range of topical interests of users in this Twitter campaign, by comparing user topical entropy ($Ent(u)$ in Eqn.~\ref{entropy}) against the user's daily average tweeting volume. We observe a large number of users with high topic entropy, indicating that most users tweet about a wide range of topics. Across both periods, we also observe a small number of outlier users with high entropy ($\ge 1$) and a high daily tweeting volume ($\ge 200$).

\subsection{Top Influential Users and Topics}
We calculate the $Inf(u)$ scores (Eqn.~\ref{userInfluence} in \S\ref{sectInfluenUser}) for all users in our dataset and identified the top 100 users for our study. For brevity, we only present the top 10 influential users and their respective categories, e.g., political parties, politicians, news/media, as shown in Table~\ref{top10user}. Other than the campaign influencers (political parties and politicians), this result highlights that there a wide range of other influential users in the top 100, including organization and personal accounts (news/media, journalists, activist groups, individuals, etc).

\setlength{\columnsep}{10pt}
\begin{wraptable}{R}{0.58\columnwidth}
\vspace{-11mm}
\caption{Top 10 Influential User (and category)}
\label{top10user}
\centering
\resizebox{0.6\columnwidth}{!} {
\begin{tabular}{C{1.5cm} p{5.5cm} p{5.5cm}} \hline
{\bf Rank}	&	{\bf Period 1}	&	{\bf Period 2}	\T \B \\ \hline
1	&	AfD\_Bund (Political Party)	&	tagesschau (News/Media)	\T \B \\
2	&	tagesschau (News/Media)	&	welt (News/Media)	\T \B \\
3	&	DefendEuropeID (Activist Group)	&	AfD (Political Party)	\T \B \\
4	&	MartinSchulz (Politician)	&	MartinSchulz (Politician)	\T \B \\
5	&	welt (News/Media)	&	SteinbachErika (Politician)	\T \B \\
6	&	Beatrix\_vStorch (Politician)	&	Einzelfallinfos	 (Political Site)	\T \B \\
7	&	Wahlrecht\_de (Election Website)	&	Beatrix\_vStorch (Politician)	\T \B \\
8	&	SteinbachErika (Politician)	&	faz\_donalphonso (Journalist)	\T \B \\
9	&	DoraBromberger (Individuals)	&	RT\_Deutsch (News/Media)	\T \B \\
10	&	LetKiser (Individuals)	&	Wahlrecht\_de (Election Website)	\T \B \\ \hline

\end{tabular}
}
\vspace{-8mm}
\end{wraptable}

Next, we examine the topical interests $\vec{I_u}$ (see \S\ref{sectUserTopic}) of the top 10 influential users. 
Unsurprisingly, we find that the news/media accounts (tagesschau, welt, RT\_Deutsch) cover a wider range of topics, while the campaign influencer accounts (AfD\_Bund, AfD, MartinSchulz, Beatrix\_vStorch, SteinbachErika) cover a more focused set of topics. Interestingly, our methodlogy was able to identify AfD\_Bund (Period 1) that was replaced by AfD (Period 2) on 03 Sep 2017 as the official Alternative for Germany account in spite of us not directly following AfD, which further demonstrate the effectiveness of our methodology.

\begin{figure} [t!]
  	\centering
    \includegraphics[width=0.37523\columnwidth, keepaspectratio=true, angle=0, trim={8mm 0mm 0mm 0mm}]{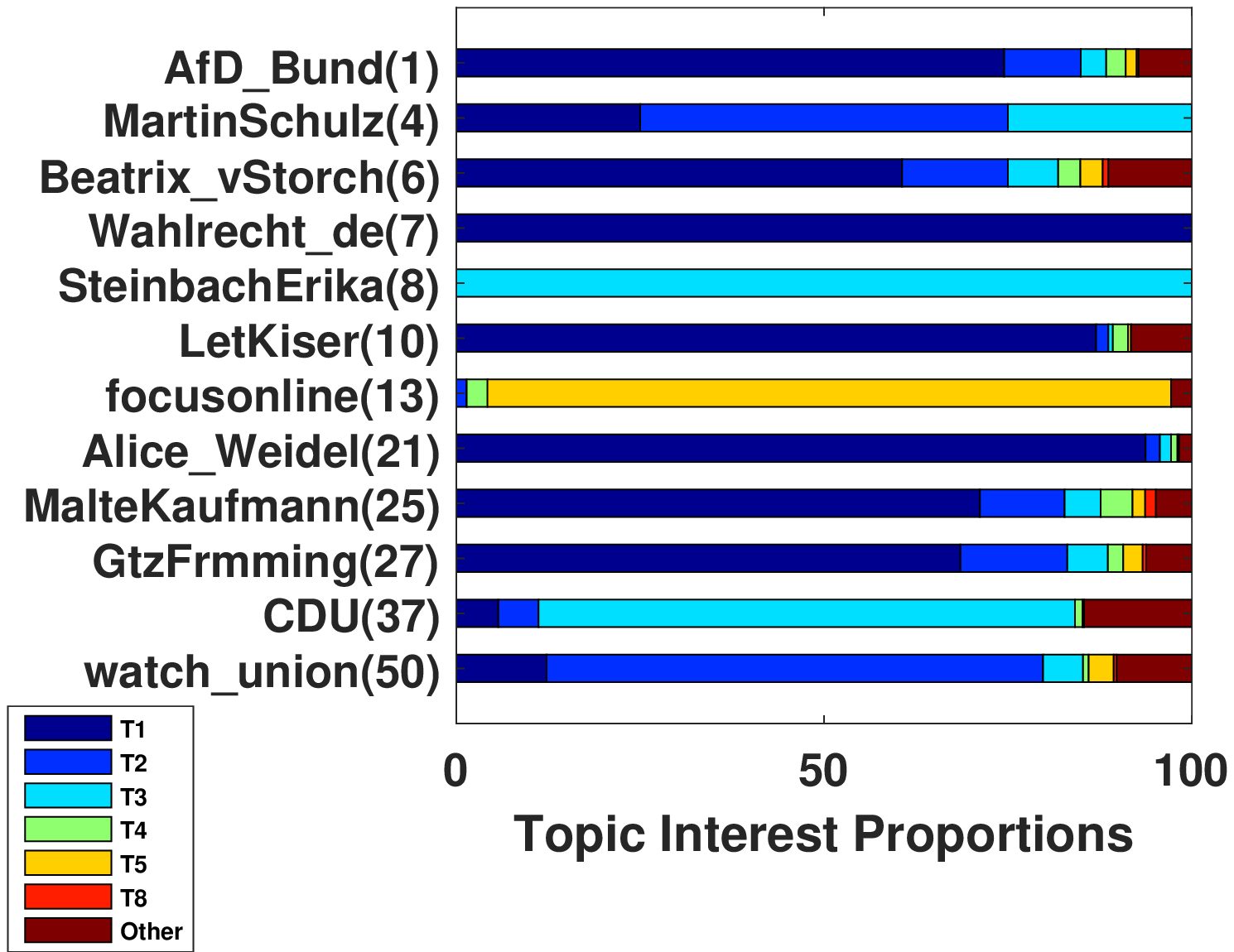}\hspace{10mm}
    \includegraphics[width=0.37523\columnwidth, keepaspectratio=true, angle=0, trim={8mm 0mm 0mm 0mm}]{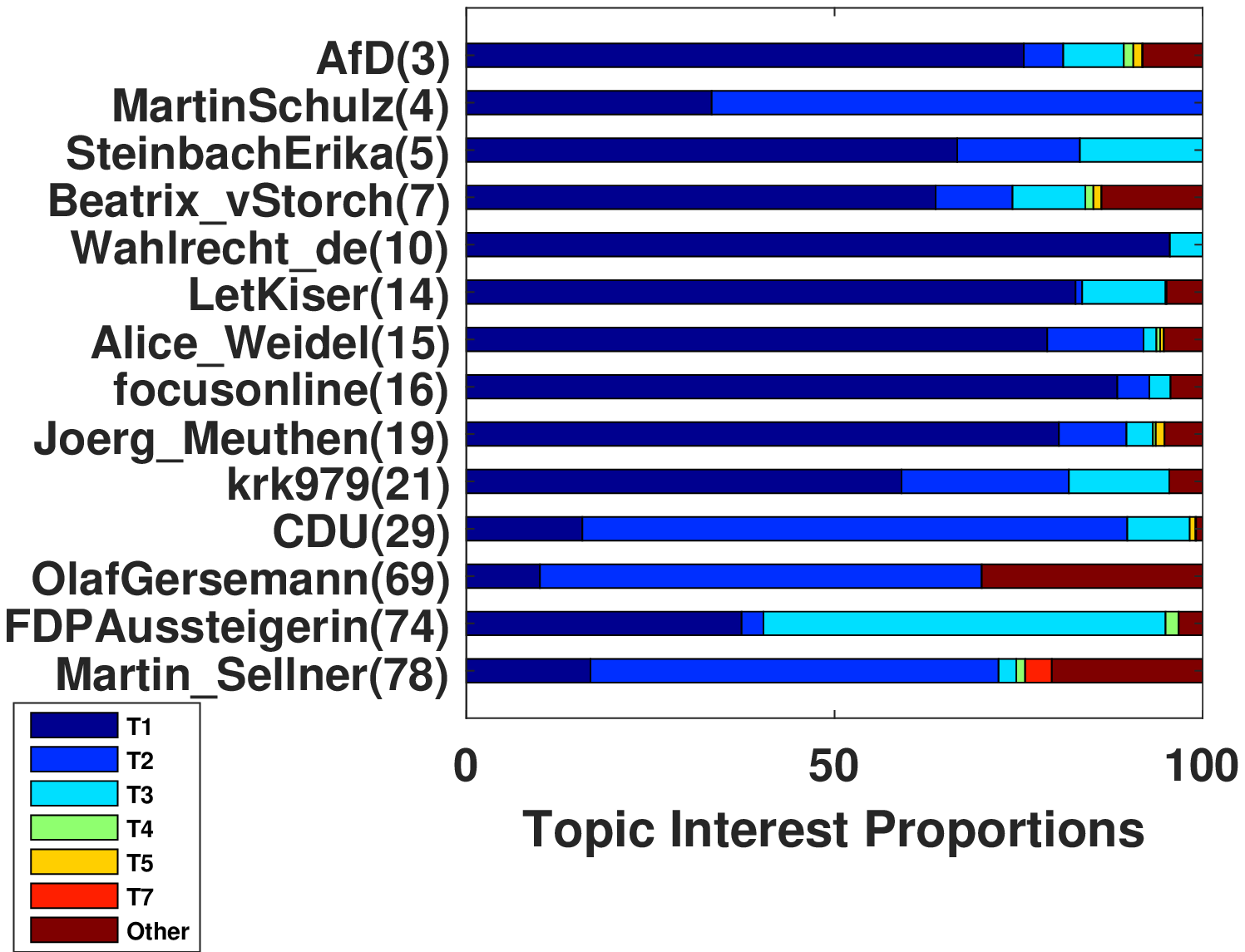}
    \caption{Topics of Top Influential Users (Politicians and Political Parties), for Period 1 (left) and Period 2 (right)}
    \label{topPolUserTopic}
\end{figure}

In this study, we are interested in Twitter campaigns and its key influencers/stakeholders, e.g., politicians/parties for elections and companies for marketing/advertising campaigns. 
Based on our list of top 100 influential users, we proceed to select a set of representative influentials by filtering out users who had more than 50\% interest in any single topic (i.e., $int^u_t > 0.5$, refer to Eqn.~\ref{topicInterest}). This selection criteria results in a list of representative influential users, as shown in Fig.~\ref{topPolUserTopic} (the number next to each influential user listed on the y-axis shows the user's influence ranking). We note that many of these users are highly representative of the key political parties and politicians in this campaign, thus validating the effectiveness of our methodology to identify influential users.

\subsection{Influential Users and their Bot Activities}

To analyse bot activity, we selected eight users that are representative of the selected topics of interest, i.e., users with more than 50\% interest in any single topic (Fig.~\ref{topPolUserTopic}). 

\vspace{-3.0425mm}
\subsubsection{Identifying Bot Activities using Unique Retweet Ratio}
\label{sectResUniRetweet}

Table~\ref{botUniqueRetweeter} shows the average unique retweet ratio $UR^p(u)$ (Eqn.~\ref{uniqueRetweet}) for the period under consideration for each of the users. The three users with the lowest unique retweet ratio are (in ascending order) AfD\_Bund, Beatrix\_vStorch, SteinbachErika for Period 1, and SteinbachErika, Beatrix\_vStorch, AfD for Period 2. 

\setlength{\columnsep}{10pt}
\begin{wraptable}{R}{0.4\columnwidth}
\vspace{-3mm}
\caption{Unique Retweeters for Influential Users, for Period 1 (left) and Period 2 (right)}
\label{botUniqueRetweeter}
\centering
\resizebox{2.3cm}{!} {
\begin{tabular}{l c} \hline
\multicolumn{2}{c}{\bf Period 1}	\T \\
{\bf User}	&	{\bf $UR^P(u)$}	\B \\ \hline
AfD\_Bund	&	0.13	\T \B \\
Beatrix\_vStorch 	&	0.15	\T \B \\
SteinbachErika 	&	0.16	\T \B \\
CDU 	&	0.29	\T \B \\
watch\_union 	&	0.3	\T \B \\
LetKiser 	&	0.42	\T \B \\
MartinSchulz 	&	0.46	\T \B \\
Wahlrecht\_de 	&	0.46	\T \B \\ \hline
\end{tabular}
}
\resizebox{2.4cm}{!} {
\begin{tabular}{l c} \hline
\multicolumn{2}{c}{\bf Period 2}	\T \\
{\bf User}	&	{\bf $UR^P(u)$}	\B \\ \hline
SteinbachErika 	&	0.12	\T \B \\
Beatrix\_vStorch 	&	0.16	\T \B \\
AfD  	&	0.18	\T \B \\
CDU 	&	0.21	\T \B \\
MartinSchulz 	&	0.49	\T \B \\
Wahlrecht\_de 	&	0.49	\T \B \\
FDPAussteigerin 	&	0.59	\T \B \\
OlafGersemann 	&	0.74	\T \B \\ \hline
\end{tabular}
}
\vspace{-6mm}
\end{wraptable}

\vspace{-3.0425mm}
\subsubsection{Identifying Bot Activities using Direct Retweeter Influence}

\begin{figure} [t!]
  	\centering
    \includegraphics[width=0.37523\columnwidth, keepaspectratio=true, angle=0, trim={5mm 0mm 0mm 0mm}]{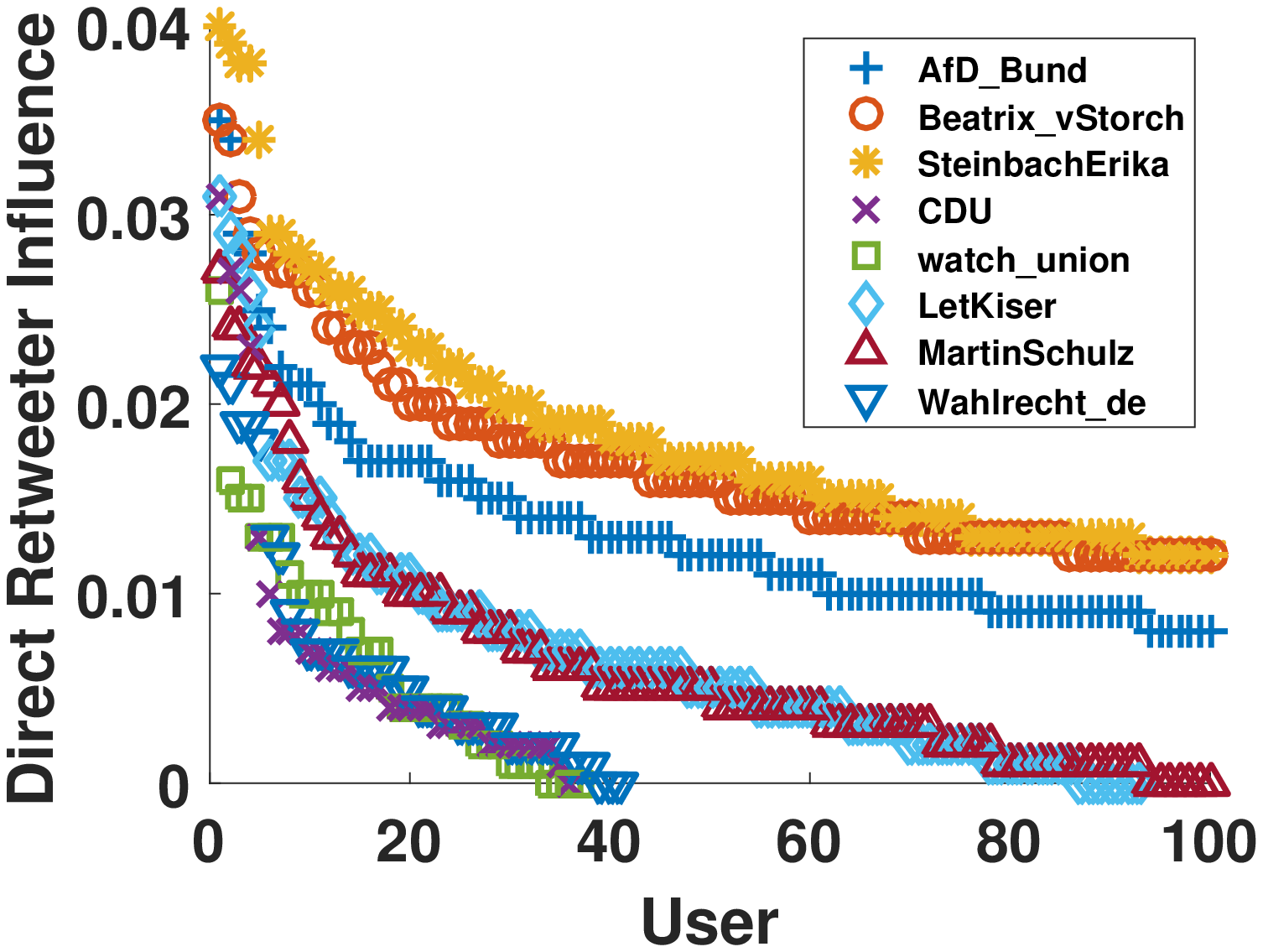}\hspace{10mm}
    \includegraphics[width=0.37523\columnwidth, keepaspectratio=true, angle=0, trim={5mm 0mm 0mm 0mm}]{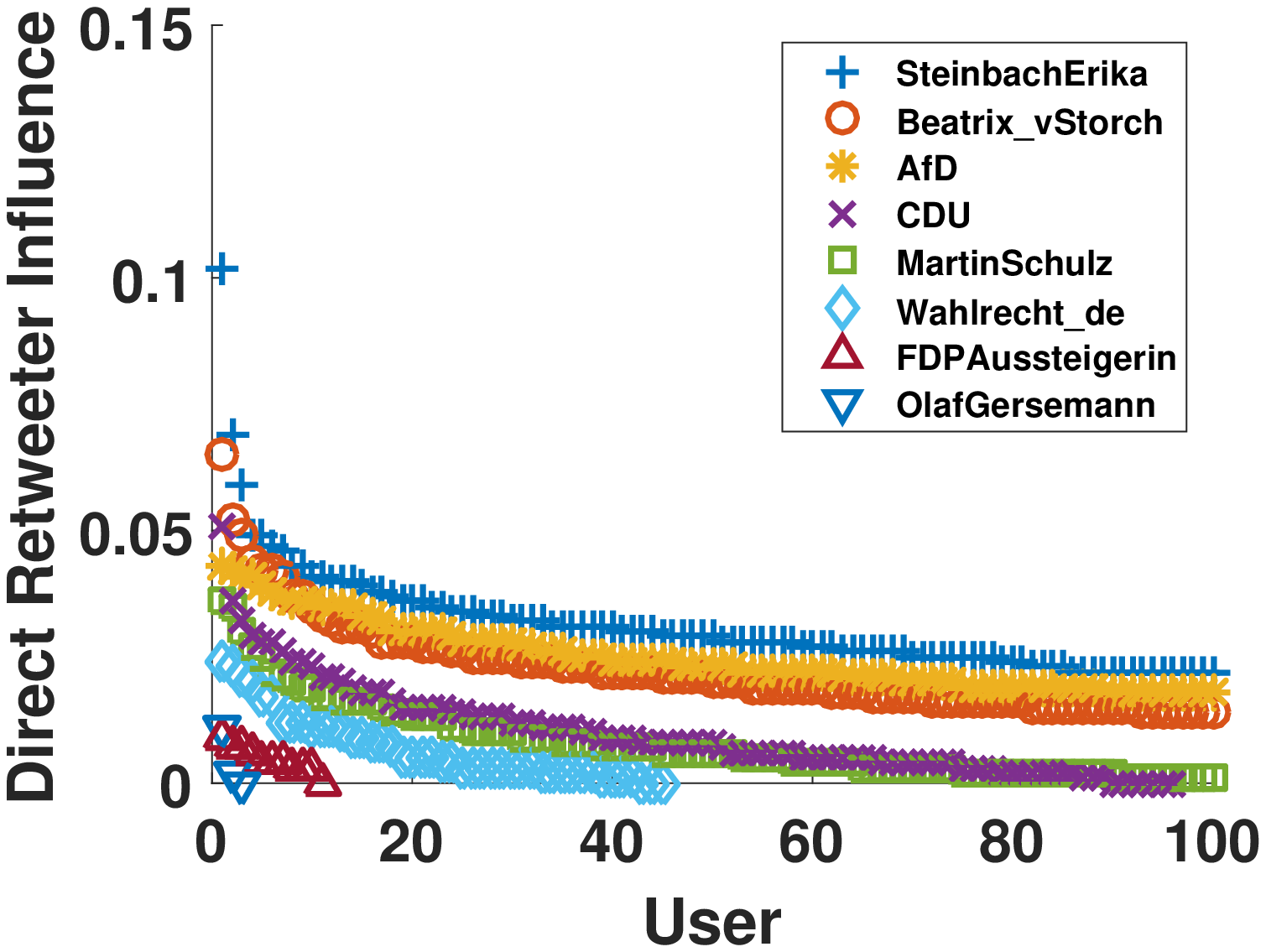}
    \caption{Direct Retweeter Influence (Transfer Entropy, refer to Eqn.~\ref{transferEntropy} for more details) for Influential Users, for Period 1 (left) and Period 2 (right).}
    \label{botTransferEntropy}
\end{figure}

Fig.~\ref{botTransferEntropy} shows the Direct Retweeter Influence $DI^p(u)$ (Eqn.~\ref{transferEntropy}) computed with time bins of 1 hour and a lag of 1, chosen as per~\cite{Steeg-www12}. Apart from having retweeters with low unique retweet ratios, the same set of campaign influencers (AfD\_Bund, Beatrix\_vStorch, SteinbachErika for Period 1, and SteinbachErika, Beatrix\_vStorch, AfD for Period 2) 
also have a large number of retweeters with high values of direct retweeter influence (Eqn.~\ref{transferEntropy}), as shown in Fig.~\ref{botTransferEntropy}. This result show that these campaign influencers are more likely to be employing bots for retweeting (compared to the other influential users), due to the high information transfer~\cite{Steeg-www12}.

\vspace{-3.0425mm}
\subsubsection{Identifying Bot Activities using Retweet Volume}

\begin{figure} [t!]
  	\centering
    \includegraphics[width=0.37523\columnwidth, keepaspectratio=true, angle=0, trim={5mm 0mm 0mm 0mm}]{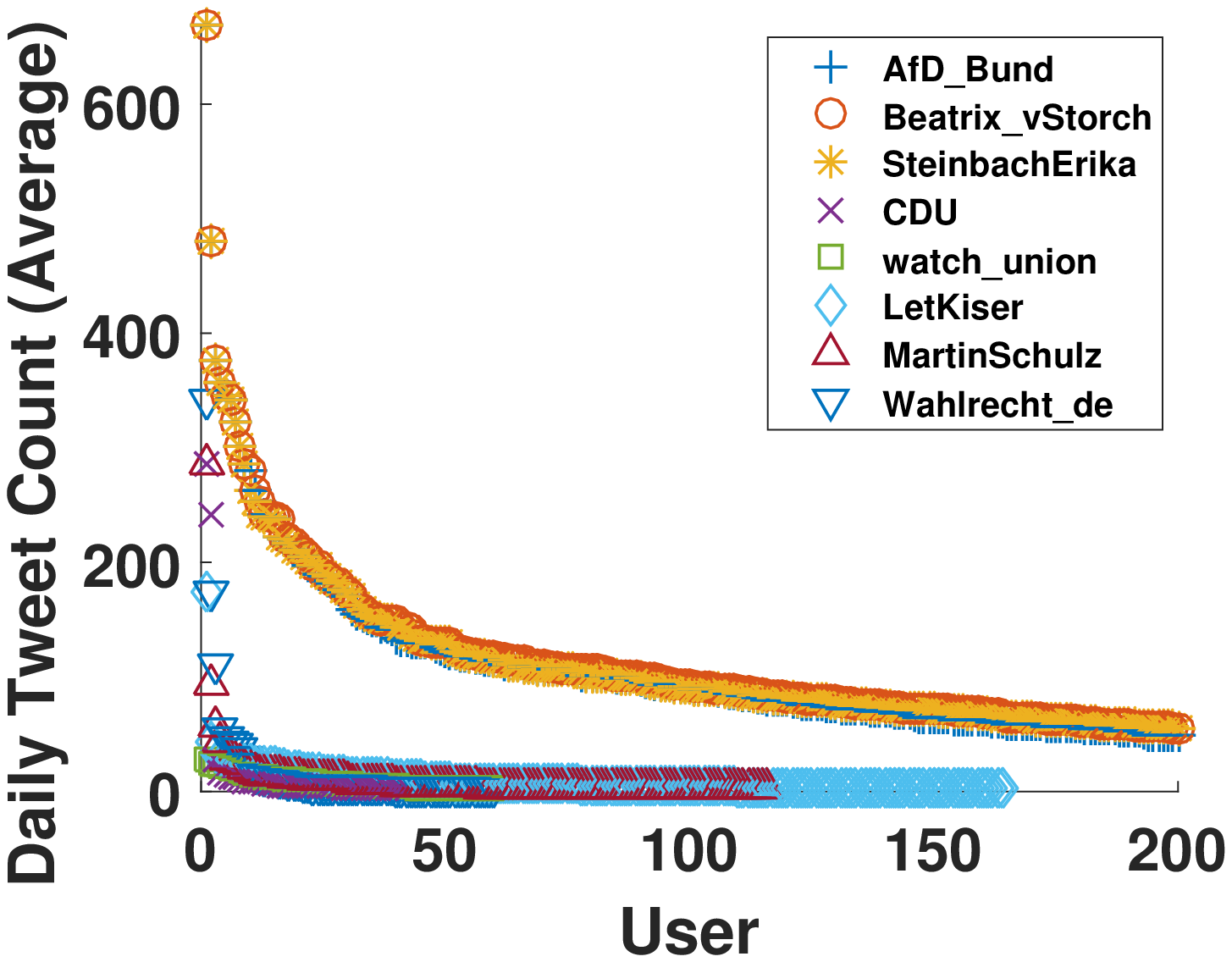}\hspace{10mm}
    \includegraphics[width=0.37523\columnwidth, keepaspectratio=true, angle=0, trim={5mm 0mm 0mm 0mm}]{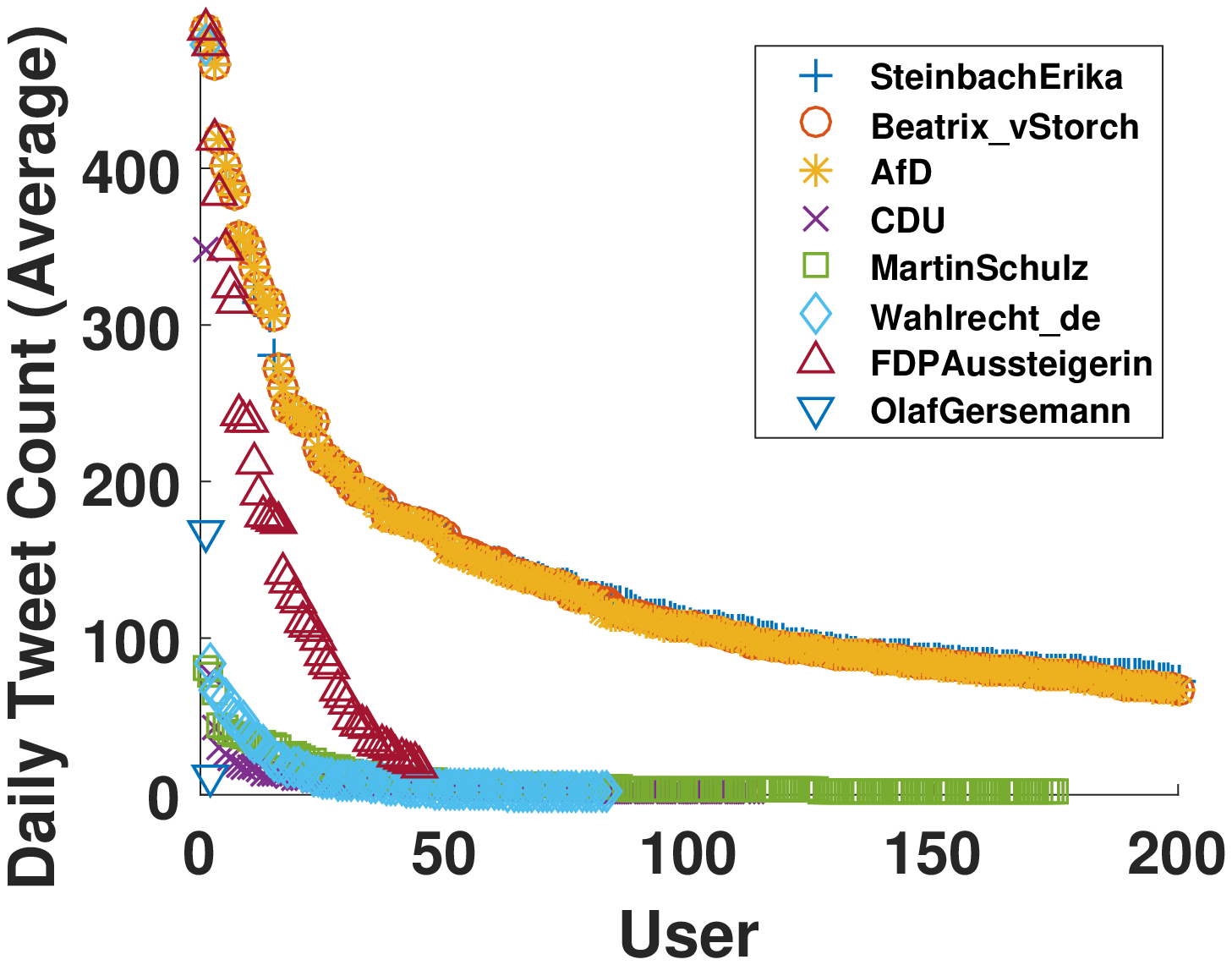}
    \caption{Retweeting Counts for Influential Users, for Period 1 (left) and Period 2 (right)}
    \label{botRetweetCounts}
\end{figure}

Fig.~\ref{botRetweetCounts} shows the average daily tweet volumes of users who retweeted the campaign influencers. From this figure, we observe that several of the retweeters of the campaign influencers display high volume of daily tweets compared to the others, e.g., AfD\_Bund, Beatrix\_vStorch and SteinbachErika in Period 1 and AfD, Beatrix\_vStorch and SteinbachErika in Period 2. 
This list of users is identical to the list of users with the lowest unique retweet ratios and highest direct retweeter influence, thus reinforcing our earlier claim that these users are more likely to be employing bots, compared to the other influential users.

\begin{figure} [t!]
  	\centering
    \includegraphics[width=0.313512\columnwidth, keepaspectratio=true, angle=0]{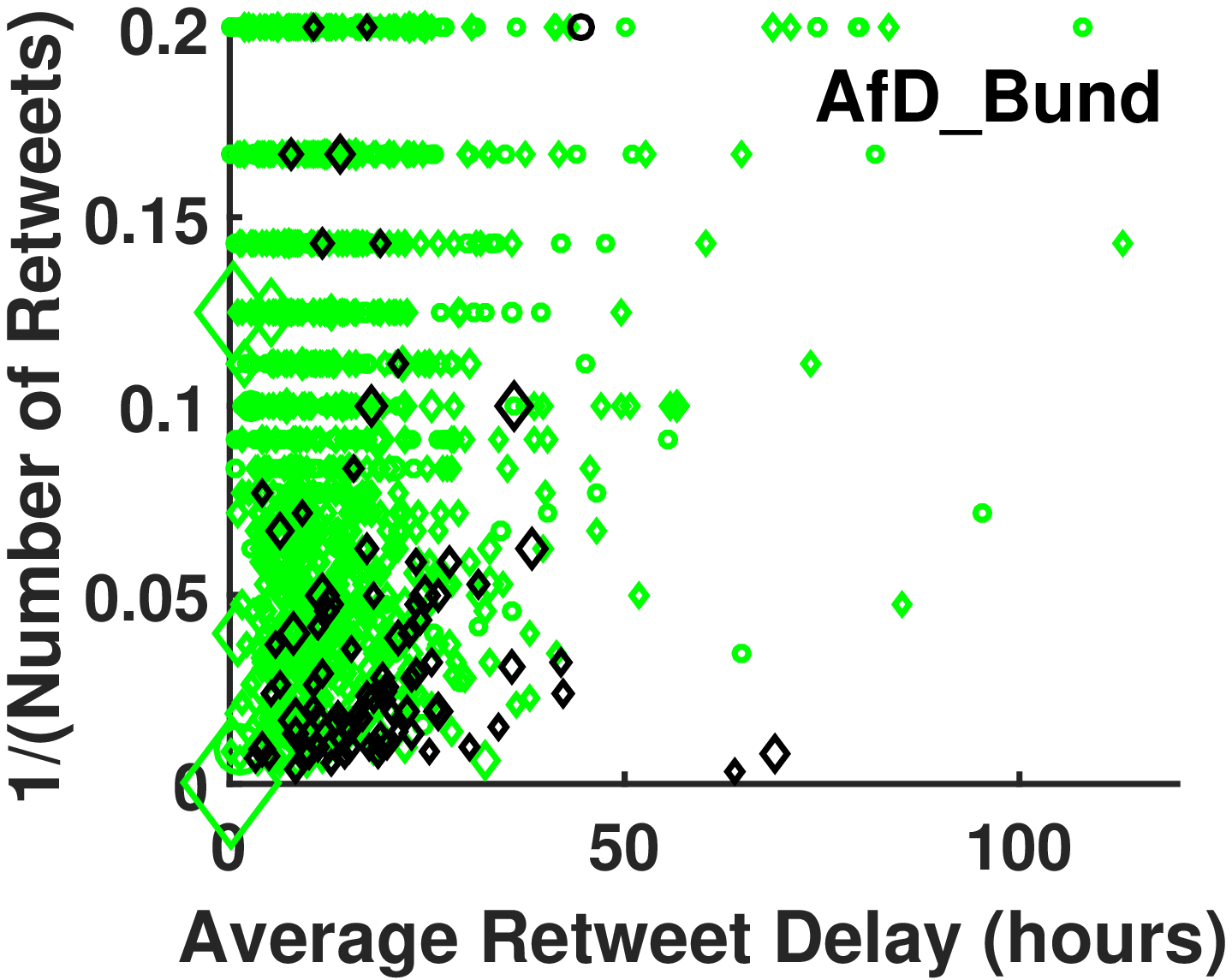}\hspace{2mm}
    \includegraphics[width=0.313512\columnwidth, keepaspectratio=true, angle=0]{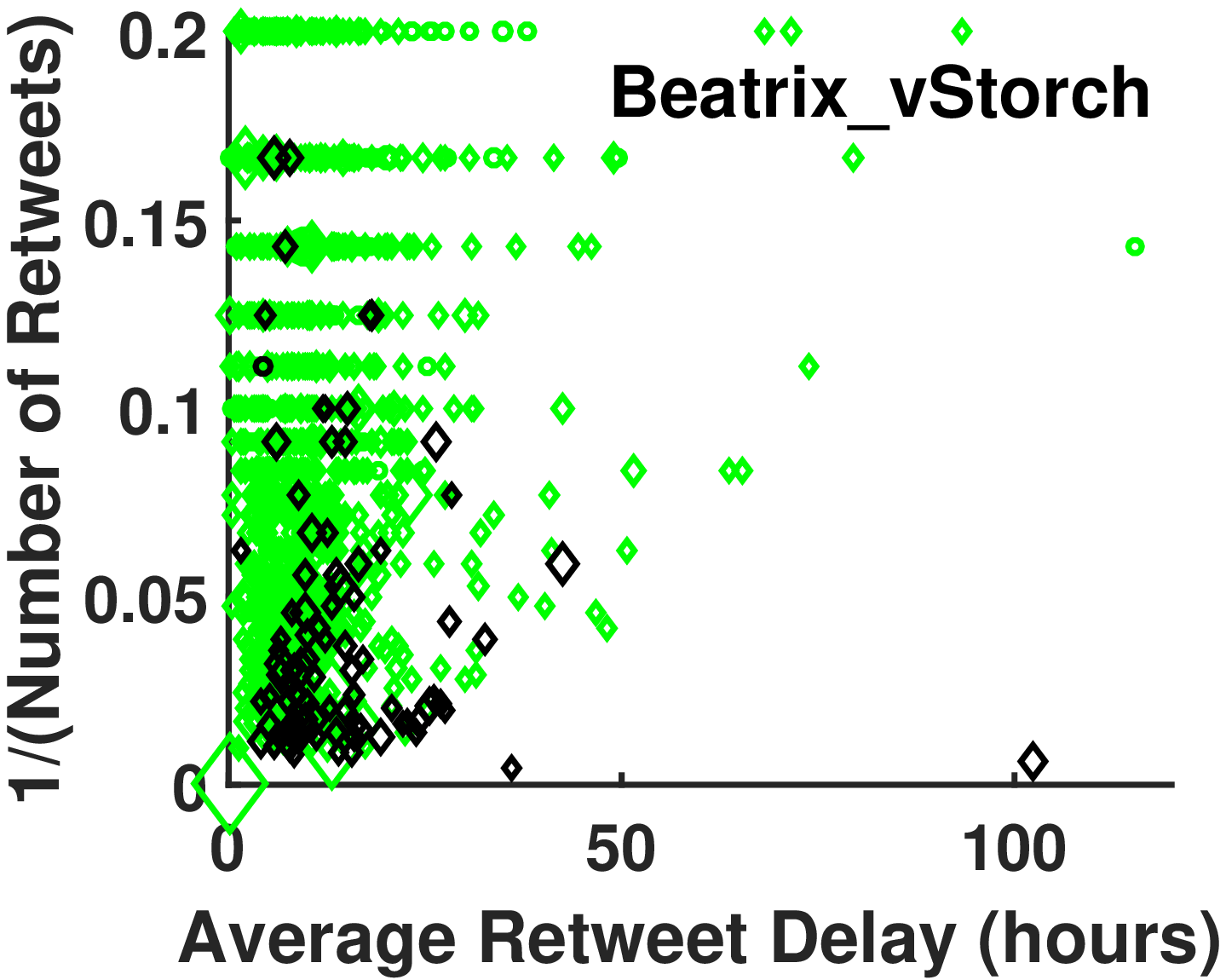}\hspace{2mm}
    \includegraphics[width=0.313512\columnwidth, keepaspectratio=true, angle=0]{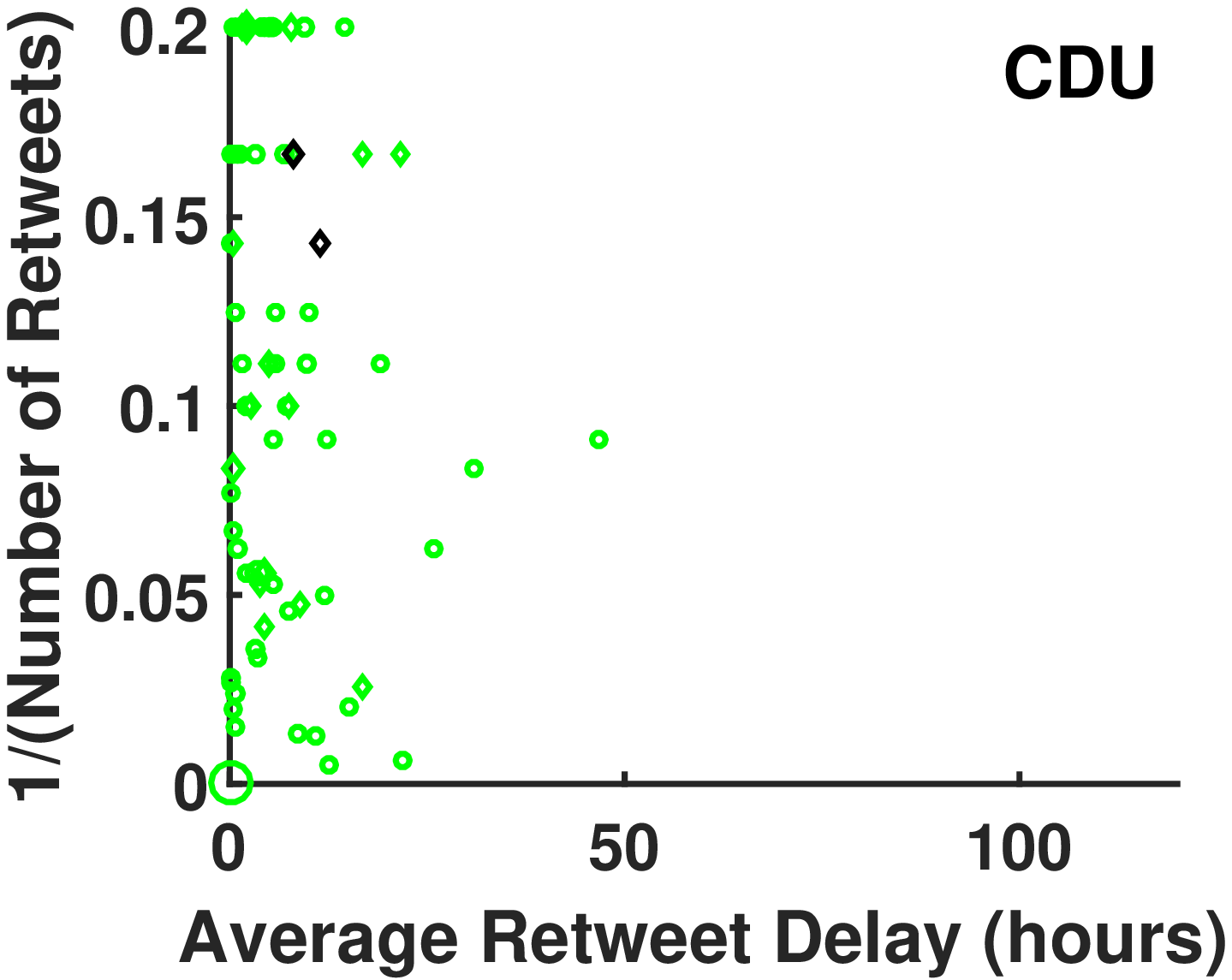}
    \caption{Visualization of Retweeters for Influential Users, for Period 1. SteinbachErika show similar trends to AfD\_Bund and Beatrix\_vStorch, while the remaining influential users are similar to CDU, but are omitted for space.}    
    \label{botRetweetVisual}
\end{figure}

\vspace{-3.0425mm}
\subsubsection{Visualization of Bot Activities}

\setlength{\columnsep}{10pt}
\begin{wraptable}{R}{0.67\columnwidth}
\vspace{-10mm}
\caption{Retweet Jaccard $RJ(u)$ (Row 9) and Pair-wise Jaccard (Rows 1 to 8) for all Influential Users, in Period 1}
\label{allJaccard}
\centering
\resizebox{0.65\columnwidth}{!} {
\begin{tabular}{c c c c c c c c c} \hline \hline
	&	AfD	&	Beatrix	&	Steinbach	&		&	watch	&	Let	&	Martin	&	Wahlrecht	\T \\
	&	\_Bund	&	\_vStorch	&	Erika	&	CDU	&	\_union	&	Kiser	&	Schulz	&	\_de	\B \\ \hline \hline
AfD\_Bund	&	1	&	0.4803	&	0.3879	&	0.0028	&	0	&	0	&	0.0016	&	0.0074	\T \B \\
Beatrix\_vStorch	&	0.4803	&	1	&	0.5023	&	0.0019	&	0	&	0	&	0.0008	&	0.0066	\T \B \\
SteinbachErika	&	0.3879	&	0.5023	&	1	&	0.0030	&	0	&	0	&	0.0026	&	0.0069	\T \B \\
CDU	&	0.0028	&	0.0019	&	0.0030	&	1	&	0	&	0	&	0.0039	&	0.0703	\T \B \\
watch\_union	&	0	&	0	&	0	&	0	&	1	&	0.0650	&	0.0350	&	0.0068	\T \B \\
LetKiser	&	0	&	0	&	0	&	0	&	0.0650	&	1	&	0.0179	&	0.0036	\T \B \\
MartinSchulz	&	0.0016	&	0.0008	&	0.0026	&	0.0039	&	0.0350	&	0.0179	&	1	&	0.0107	\T \B \\
Wahlrecht\_de	&	0.0074	&	0.0066	&	0.0069	&	0.0703	&	0.0068	&	0.0036	&	0.0107	&	1	\T \B \\ \hline
$RJ(u)$	&	{\bf 0.1257}	&	{\bf 0.1417}	&	{\bf 0.1290}	&	{\bf 0.0117}	&	{\bf 0.0153}	&	{\bf 0.0124}	&	{\bf 0.0104}	&	{\bf 0.0160	}\T \B \\ \hline \hline 
\end{tabular}
}
\vspace{-5mm}
\end{wraptable}

Fig.~\ref{botRetweetVisual} shows a visualization of the retweeters of the eight influential users in Period 1. In this figure, the sizes of individual points reflect the influence $Inf(u)$ of the retweeter, as defined in Eqn.~\ref{userInfluence}, and at the origin is the influential user itself. The shape of the points indicates the number of influential users retweeted, where circles represent retweeters who retweeted only one of the eight influential users while diamonds denote those who retweeted more than one of the identified influential users. The colour of the points denote the tweeting volume where black indicates retweeters that posted more than 100 tweets and green indicates otherwise. From these plots, we can see that AfD\_Bund, Beatrix\_vStorch and SteinbachErika 
have many retweeters that exhibit bot-like behaviour (high tweeting volumes and retweeting many of the influential users tweets) for Period 1. The plots for Period 2 showed similar patterns, but are not presented here due to space.

\vspace{-3.0425mm}
\subsubsection{Common Retweeters across Campaign Influencers}

Table~\ref{allJaccard} shows the Retweet Jaccard $RJ(u)$ (Eqn.~\ref{retweetJaccard} from \S\ref{sectBotActivities}) for all influential users (last row) and the Jaccard similarity for all pair-wise user combinations (first 7 rows). From this table, we can see that Beatrix\_vStorch, AfD\_Bund and SteinbachErika 
have high $RJ(u)$ scores as well as high pair-wise Jaccard similarity with each other. This result indicates that there is a large overlap of common retweeters among the three users. We observe similar trends for Period 2.

In contrast, the remaining four influential users have extremely small values for Retweet Jaccard $RJ(u)$, which are one tenth that of Beatrix\_vStorch, AfD\_Bund and SteinbachErika. 
In terms of pair-wise Jacacrd similarity, these four influential users also display very small values comapred to other users, and in many cases a pair-wise Jacacrd similarity of 0, which indicates no common retweeter among two users.

\vspace{-3.0425mm}
\subsubsection{Identities of Common Retweeters}

Next, we perform a qualitative analysis on the individual retweeters for Beatrix\_vStorch, AfD\_Bund and SteinbachErika, focusing on their top 5 retweeters (in terms of retweeting volume) and the rank at which these retweeters appear for the other users if they were not already in the top 5. The results support our earlier observation that all three influential users share a high overlap in terms of their common retweeters.

\section{Conclusion}
\label{sectConclusion}

We proposed a novel methodology for analyzing Twitter campaigns, focusing on various critical tasks to holistically and better understanding campaigns, such as campaign topic detection, filtering important topics, identifying influential users, studying user interactions and detecting bot-like activities. More specifically, we developed numerous algorithms and approaches for each task, combining techniques from various problem domains and proposing various useful measures. 
We demonstrate a use-case on the 2017 German federal election and show our methodology effectively identifies the important topics, influential users and their campaign strategies, including possible bot-involved dissemination techniques.

Although we demonstrated on an election-type campaign, this methodology is also generalizable to any Twitter campaign by providing the appropriate set of hashtags and users, e.g., \#Rio2016, \#olympian, @OlympicCh, @Rio2016\_en for an Olympic advertising campaign.



\vspace{2.5mm}
{\small
\noindent {\em \bf Acknowledgments.} 
This research is supported by Defence Science and Technology. The authors thank Aram Galstyan 
for his useful comments on this work.
}

\bibliographystyle{splncs04}
\bibliography{electionStudyShort}  

\begin{thebibliography}{10}
\providecommand{\url}[1]{\texttt{#1}}
\providecommand{\urlprefix}{URL }
\providecommand{\doi}[1]{https://doi.org/#1}

\bibitem{An-sci11}
An, X.Y., Wu, Q.Q.: Co-word analysis of the trends in stem cells field based on
  subject heading weighting. Scientometrics  \textbf{88}(1) (2011)

\bibitem{blondel-jsm08}
Blondel, V.D., Guillaume, J.L., Lambiotte, R., Lefebvre, E.: Fast unfolding of
  communities in large networks. J.\ of Statistical Mechanics
  \textbf{2008}(10),  P10008 (2008)

\bibitem{Boutet-icwsm12}
Boutet, A., Kim, H., Yoneki, E.: What's in your tweets? i know who you
  supported in the uk 2010 general election. In: ICWSM'12 (2012)

\bibitem{Brin-cn12}
Brin, S., Page, L.: The anatomy of a large-scale hypertextual web search
  engine. Computer Networks  \textbf{56}(18) (2012)

\bibitem{Conover-passat11}
Conover, M.D., Goncalves, B., Ratkiewicz, J., Flammini, A., Menczer, F.:
  Predicting the political alignment of twitter users. In:
  PASSAT'11/SocialCom'11 (2011)

\bibitem{Davis-www16}
Davis, C.A., Varol, O., Ferrara, E., Flammini, A., Menczer, F.: Botornot: A
  system to evaluate social bots. In: WWW'16 (2016)

\bibitem{Hall-emnlp08}
Hall, D., Jurafsky, D., Manning, C.D.: Studying the history of ideas using
  topic models. In: EMNLP'08 (2008)

\bibitem{Hirsch-pnas05}
Hirsch, J.E.: An index to quantify an individual's scientific research output.
  PNAS  \textbf{102}(46) (2005)

\bibitem{Kwak-www10}
Kwak, H., et~al.: What is twitter, a social network or a news media? In: WWW'10

\bibitem{lim-wias16}
Lim, K.H., Datta, A.: An interaction-based approach to detecting highly
  interactive twitter communities using tweeting links. Web Intelligence
  \textbf{14}(1) (2016)

\bibitem{lim-ecml18}
Lim, K.H., Jayasekara, S., Karunasekera, S., Harwood, A., Falzon, L., Dunn, J.,
  Burgess, G.: {RAPID: Real-time Analytics Platform for Interactive Data
  Mining}. In: {ECML-PKDD'18} (2018)

\bibitem{lim-bigdata17}
Lim, K.H., Karunasekera, S., Harwood, A.: Clustop: A clustering-based topic
  modelling algorithm for twitter using word networks. In: BigData'17 (2017)

\bibitem{nyt-election16}
{New York Times}: Internet (2016),
  https://www.nytimes.com/2016/11/09/technology
  /for-election-day-chatter-twitter-ruled-social-media.html

\bibitem{Prasetyo-ht15}
Prasetyo, N.D., Hauff, C.: Twitter-based election prediction in the developing
  world. In: HT'15 (2015)

\bibitem{Riquelme-ipm16}
Riquelme, F., Gonzalez-Cantergiani, P.: Measuring user influence on twitter: A
  survey. Information processing \& management  \textbf{52}(5) (2016)

\bibitem{Steeg-www12}
Steeg, G.V., Galstyan, A.: Information transfer in social media. In: WWW'12

\bibitem{Tumasjan-icwsm10}
Tumasjan, A., Sprenger, T.O., {et al.}: Predicting elections with twitter: What
  140 characters reveal about political sentiment. In: ICWSM'10 (2010)

\bibitem{Varol-icwsm17}
Varol, O., Ferrara, E., Davis, C.A., Menczer, F., Flammini, A.: Online
  human-bot interactions: Detection, estimation, and characterization. In:
  ICWSM'17 (2017)

\bibitem{Weng-wsdm10}
Weng, J., Lim, E.P., Jiang, J., He, Q.: Twitterrank: finding topic-sensitive
  influential twitterers. In: WSDM'10 (2010)

\bibitem{Xiao-jwe14}
Xiao, F., et~al.: Finding news-topic oriented influential twitter users based
  on topic related hashtag community detection. J.\ of Web Engineering
  \textbf{13}(5-6) (2014)

\end{thebibliography}

\end{document}